\documentstyle[aps,prb,multicol]{revtex}

\begin{document}
\input{epsf}
\draft
\preprint{}
\title{Single hole transistor in a p-Si/SiGe quantum well}
\author{U. D\"otsch$^{1}$, U. Gennser$^{1}$, T. Heinzel$^{2}$, S. 
L\"uscher$^{2}$, C. David$^{1}$, G. Dehlinger$^{1}$,  
 D. Gr\"utzmacher$^{1}$,  and K. Ensslin$^{2}$}
\address{$^{1}$Paul Scherrer Institute, 5232 Villigen PSI, 
Switzerland\\
$^{2}$Solid State Physics Laboratory, ETH Z\"{u}rich, 8093
Z\"{u}rich,  Switzerland\\}

\date{\today}
\maketitle
\begin{abstract}
A single hole transistor is patterned in a p-Si/SiGe quantum well by 
applying voltages to nanostructured top gate electrodes. Gating is 
achieved by oxidizing the etched semiconductor surface and the mesa walls before 
evaporation of the top gates. Pronounced Coulomb blockade effects are 
observed at small coupling of the transistor island to source 
and drain.
\end{abstract}
\begin{multicols}{2}
\narrowtext
Transport through small metallic islands, defined in semiconductor 
heterostructures, has been of considerable interest recently.\cite{Kouwenhoven97} 
When the island is only weakly coupled to reservoirs via 
tunnel barriers, the Coulomb blockade (CB) effect determines the carrier 
transport. In such islands, the interplay between 
interactions and size quantization can be 
studied. A large number of experiments have investigated the 
properties of single electron transistors made out of conventional 
metals, and of high mobility, low density two-dimensional 
electron gases, defined in mostly Ga[Al]As or Si, among other 
materials.\cite{Kouwenhoven97} Single hole transistors out of  
p-Si/SiGe quantum wells are of particular interest for several reasons: 
(i) the material has a large g-factor of $g_{eff}$ = 6.7 and a large effective 
mass ($m^{*}$ = 0.25), (ii) the interaction parameter $r_{S}$, defined as the ratio 
between the Coulomb energy of two holes at their average separation, 
and the Fermi energy, is very large, i.e., $r_{S}\approx 4$ for 
typical carrier densities, and (iii) the properties of the quantum well 
are strongly influenced by biaxial strain, due to the lattice mismatch at 
the Si/SiGe interface.\\
However, p-SiGe has been notoriously difficult to gate, due to leakage 
currents across the Schottky barrier.\cite{Coleridge96,Madhavi2000} 
Recently, progress has been reported in this respect. Emeleus et al. \cite{Emeleus97} 
have tuned the hole density in a p-Si/SiGe quantum well using a homogenous back 
gate. This method, however, cannot be applied for the production of 
nanostructured gates. Homogeneous top-gating of this material has been reported 
as well.\cite{Sadeghzadeh2000} In those samples, the 
two-dimensional hole gas (2DHG) is buried very deeply below the surface, which limits 
the pattern transfer into the 2DHG. To our knowledge, the fabrication of any kind of 
gated nanostructure in p-Si/SiGe has not been reported yet. It should 
be noted that small islands and wires have been patterned out of p-$Si/Si_{1-x}Ge_{x}/Si$ 
resonant tunneling structures by etching.\cite{Akyuz98} Furthermore, 
single hole charging effects have been observed in p-Si\cite{Leobandung1995} and in 
two-terminal,$\delta$-doped SiGe devices\cite{Paul1994}. \\

Here, we present the fabrication and characterization of a gated single 
hole transistor in a p-Si/SiGe quantum well. The gate leakage is 
reduced by oxidizing the etched surface before the definition of the 
top gates. At low temperatures, we observe, clear single hole tunneling and CB oscillations.\\
 The samples have been grown by molecular beam 
epitaxy, and contain a 200{\AA} $\rm{Si_{0.85}Ge_{0.15}}$ layer on 
top of an undoped Si substrate, a 150{\AA} B-doped Si layer with a setback of 
180{\AA} from the well, and a 200{\AA} undoped Si cap. The $Si_{0.85}Ge_{0.15}$ layer forms a
triangular potential well for the two-dimensional hole gas (Fig. 1a). Due to 
the lattice mismatch between Si and $Si_{0.85}Ge_{0.15}$, the 
heavy hole ($m_J$ = $\pm$3/2) potential is split from the light hole ($m_J$ =
$\pm$1/2) potential, and ensures that the lowest occupied bound state 
has heavy hole character. The hole density of the ungated 2DHG is 
$p_{2D}$ = 3.6 $\times$ 10$^{15}$m$^{-2}$, and an elastic mean free 
path of $l_{e}$ = 100 nm is obtained from magnetoresistivity 
measurements.\\
\begin{figure}
     \centerline{\epsfxsize=7.0cm \epsfbox{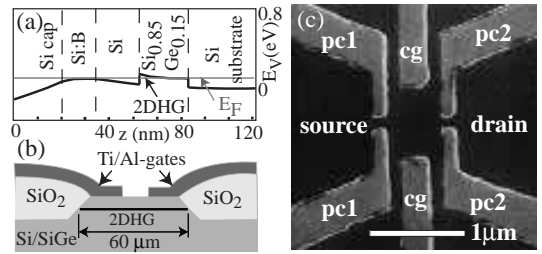}}	
       \caption[Fig. 1] {(a) Top of the valence band of the Si/SiGe 
       quantum well. The hole gas resides at the 
       Si/Si$_{0.85}$Ge$_{0.15}$ interface 53 nm below the surface. (b) 
       Schematic cross section through the processed heterostructure. 
       In order to reduce leakage currents, the mesa edges are separated 
       from the top gate by a SiO$_{2}$ layer. (c) Scanning electron micrograph of the 
       gate geometry. The Ti/Al gates appear as bright areas. }
    \end{figure}
Tests have indicated that in our samples, large leakage currents between the top gate and 
the 2DHG can flow across the mesa edge. Therefore, we have deposited a 
layer of SiO$_{2}$ by thermal evaporation right after the reactive 
ion etch that defines the mesa, i.e., with the photoresist still on the 
mesa. We have used this technique in earlier work for the fabrication 
of homogeneous gates, Fig. 1b. \cite{Senz2000} After lift-off of the oxide, the
top gate (10 nm Ti and 20 nm Al) was patterned by electron beam lithography and 
thermal evaporation. This sequence of technology steps reduces leakage currents in our samples 
significantly.\\
The geometry of the gates is crucial. While deposition of a \textit{homogeneous} 
Ti/Al gate reduces $p_{2D}$ by roughly a factor of two due to pinning effects, we 
find that for grounded gates in the sub-micron regime, the 2DHG is depleted 
underneath, with a lateral depletion length of roughly 100 nm. This depletion is 
probably caused by strain and pinning on defects, which complicates a simulation of 
the device. Preliminary experimental studies on split gates with various openings $w$ 
between the two electrodes were used to develop the gate geometry. For 
$w <$ 100 nm, we were 
unable to open the channel between the split gates without simultaneously populating the 
regions below the gates. For $w >$ 250 nm, the channels could not be pinched off. Also, the 
pinch-off voltage of split gates with identical geometry can be quite different.\\ 
In Fig. 1c, the island geometry is shown. Three split gate electrodes (point contact gates 
pc1, pc2, as well as the center gates cg) define the dot with a lithographic size of 500 nm 
$\times$ 800 nm. For pc1 and pc2, $w$ = 150 nm was chosen.\\
The measurements have been carried out in a $^{3}He/^{4}He$ dilution refrigerator with a base 
temperature of 90 mK. DC voltages are applied between source and drain, and the current is 
measured with a resolution of $\approx$ 500 fA.\\
\begin{figure}
    \centerline{\epsfxsize=7.0cm \epsfbox{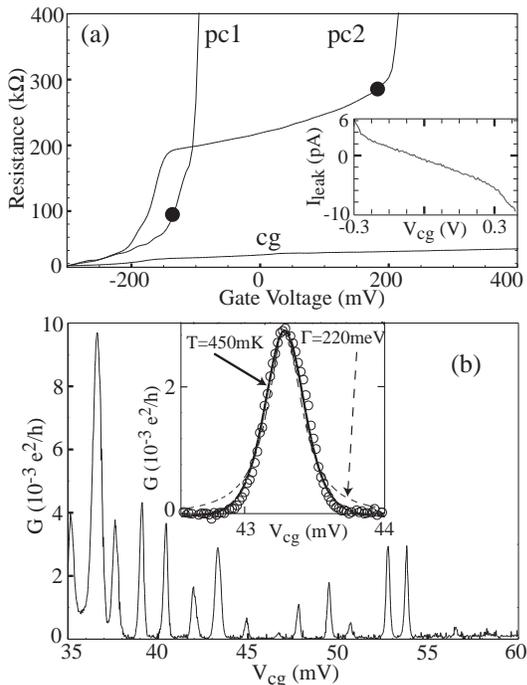}}	
       \caption[Fig. 2] {(a) Resistance of the 3 split gates as a function of the 
       gate voltages $ V_{pc1}, V_{pc2}$, and $V_{cg}$, respectively. The circles denote 
       the working points of gates pc1 and pc2 for all subsequent measurements. The inset 
       shows the leakage current of gate cg as a function of $V_{cg}$, which is typical 
       for all gates. (b) Conductance G through the island as a function of $V_{cg}$, 
       showing CB oscillations. Inset: blow-up of the peak at $V_{cg}$ = 43 mV (open circles),
       fitted to a Lorentzian (dashed line), and to a thermally broadened CB resonance 
       (solid line), see text.}
    \end{figure}
We have studied two different quantum dots of identical geometry and performed several 
cooldowns, all leading to similar results. Here, we focus on the measurements from one 
device, taken in a single cooldown.\\
\begin{figure}
     \centerline{\epsfxsize=7.5cm \epsfbox{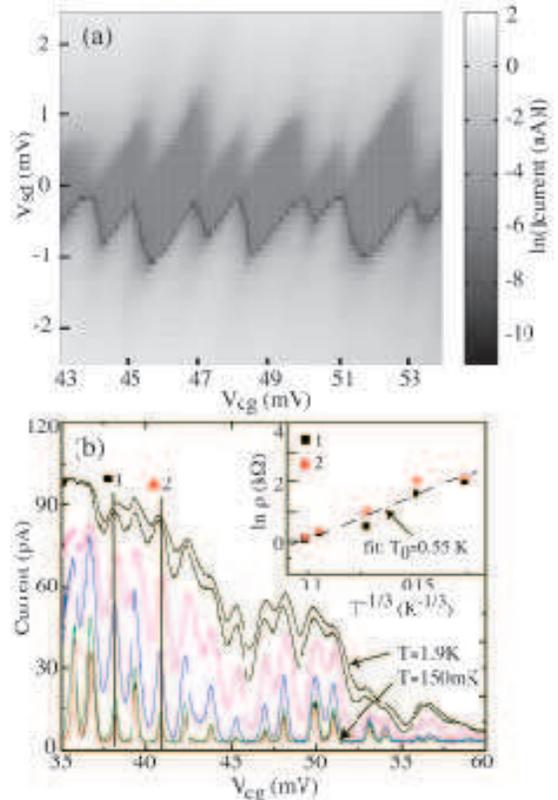}}	
       \caption[Fig. 3] {(a) The Coulomb diamonds in a logarithmic gray scale plot. A possible
       origin of the modulation is discussed in the text. (b) Temperature dependence of the CB resonances. 
       Besides a thermal smearing, a striking overall increase of G is observed as T 
       increases. The inset shows ln($\rho$) as a function of T$^{-1/3}$ at gate voltages 
       1 and 2, including a fit of the full squares. The approximately linear behavior 
       indicates thermally activated transport through the island (see text). 
      }
    \end{figure}
Split gates pc1 and pc2 are used to adjust the coupling of the island to source and drain 
(Fig. 2a). Due to the small elastic mean free path, conductance quantization is not observed. 
The leakage currents measured at the center gates is shown in the inset of Fig. 2a. At 
sufficiently large split gate voltages $V_{pc1}$ and $V_{pc2}$, CB oscillations are observed 
as a function of the center gate voltage $V_{cg}$ (Fig. 2b).\\ 
The measured CB resonances can be fitted well to the expression describing a thermally smeared 
CB peak in the multi-level transport regime, i.e. $G(V_{cg},T_{h})=G_{max}cosh^{-2}
\lbrack 
\frac{\eta(V_{cg}-V_{max})}{2.5k_{B}T_{h}}\rbrack$.\cite{Kouwenhoven97} Here, $G_{max}$ and $V_{max}$ 
denote the amplitude and the position of the CB resonance, respectively. The lever arm 
$\eta$ translates $V_{cg}$ into energy and is obtained from measurements of the CB diamonds 
(Fig. 3a). We find $\eta$ = 0.68 eV/V. As fit parameter, we obtain hole temperatures 
of $T_{h}$ = 390 mK to 570 mK, depending on the peak. A Lorentzian, on the other hand, 
which is appropriate for homogeneously broadened resonances, fits significantly worse 
(inset in Fig. 2b). The full widths at half maximum (FWHMs) thus 
fluctuate strongly from peak to peak, and furthermore saturate for bath 
temperatures below 400 mK, as well as for source-drain bias voltages below V$_{sd}$ = 
120 $\mu$V. We conclude that neither the electron temperature, nor V$_{sd}$ 
determine the FWHM in this regime (T $<$ 400 mK and 
V$_{sd}$ $<$ 120 $\mu$V). Furthermore, we observe a pronounced pair modulation of the peak 
separations in this regime. Qualitatively similar modulations are 
observed in both devices and in all cooldowns, although the detailed 
structure varies strongly. A measurement of the Coulomb diamonds, 
i.e., the current as a function of both $V_{sd}$ and 
$V_{cg}$ is shown in Fig. 3a. The stability of the diamonds 
shows that there is little telegraph noise and switching at frequencies below 1Hz, the 
bandwidth of our setup. However, noise of higher frequencies may be 
substantial; and gate voltage dependent charge rearrangements in the substrate may be present 
and could determine the FWHM of the CB peaks.\\
In Fig. 3b, we show the temperature dependence of the CB 
oscillations. Temperature dependent measurements support the 
interpretation in terms of a strongly disordered dot (Fig. 3b). 
As the temperature increases, not only the CB 
oscillations are smeared out, but also a striking overall increase in 
$G$ is observed, which indicates thermally enhanced transport 
through the island. Since the conductance of the individual tunnel 
barriers, measured with positive voltages applied to the remaining 
gates, does not depend on temperature between 90 mK and 2K,
the observed overall increase of $G$ is entirely due to the island. Several scenarios could 
lead to such a behavior, such as Variable-Range Hopping\cite{Mott1949} 
or Efros-Shklovskii hopping\cite{Efros1975}. Reasonable agreement 
with Variable-Range Hopping (i.e., $\rho\propto exp\lbrack(T_{0}/T)^{1/3}\rbrack$) is  found, and we can estimate the 
characteristic temperature from fits to $T_{0}\approx 0.55K$. This 
result depends only weakly on $V_{cg}$. We cannot, 
however, distinguish between Variable-Range hopping and 
Efros-Schklovskii hopping, due to the limited range of temperatures.\\
 Similar modulations of Coulomb diamonds have been observed in 
double dot systems for weak coupling between the dots.\cite{Crouch97} 
Possibly, the modulation is thus an indication that at 
$V_{cg} \approx$ 40 mV,
the island splits up into two smaller, weakly coupled puddles, due to disorder. 
The modulated diamonds then reflect the formation of a 
``molecular state'', or of a capacitance between the two puddles.
A different number of puddles would result in a different modulation period. Since 
we can not tune the two puddles independently nor the coupling between 
them, we are unable to extract the puddle sizes and the interdot 
conductance. Nevertheless, this interpretation allows us to quantify 
the island in terms of capacitances. Within this interpretation, the 
\textit{average} CB period in Fig. 3a
corresponds to $e/C_{g}$, where e is the elementary charge and 
$C_{g}$ the capacitance between the total island and the center gates. The 
average distance between the diamond corners along the 
$V_{b}$-direction corresponds to twice the charging energy 
$E_{C}=e^{2}/C_{\Sigma}$ of the island: $\delta V_{sd}$ = $2e/C_{\Sigma}$. 
Here, $C_{\Sigma}$ denotes the total capacitance of the island. From Fig. 
3a, we find $C_{\Sigma}$ = 170 aF, which corresponds to a single electronic island 
radius of r$\approx$ 200 nm. Hence, the level spacing inside the dot 
can be estimated as $\Delta \approx$ 8 $\mu$eV, which justifies our 
assumption of multilevel transport used above.\\

In summary, we have demonstrated that top gate defined nanostructures 
can be fabricated and operated on p-SiGe. The leakage currents can be 
significantly reduced by oxidizing the mesa edges before top gate 
deposition. We have observed Coulomb blockade on a top gate defined 
island in p-Si/SiGe. The island is strongly disordered, which manifests 
itself in thermally enhanced transport as well as in a possible puddle formation 
at small hole densities.\\
The authors would like to thank S. Graf for technical assistance. Financial support by the Swiss National Science Foundation and MINAST is 
gratefully acknowledged.

\end{multicols}

\end{document}